\documentclass[aps,prb, groupedaddress,twocolumn]{revtex4-1}
\usepackage{times,xspace}
\usepackage{amsfonts}
\usepackage{amsmath}
\usepackage{amssymb}
\usepackage[final]{graphicx}
\begin{document}
\title{Reply to ``Comment on 'Vortex-assisted photon counts and their magnetic field dependence in single-photon superconducting detectors'"}
\author{L.~N.~Bulaevskii, Matthias Graf$^1$ and V.G. Kogan$^2$} 
\affiliation{$^1$Los Alamos National Laboratory, Los Alamos, New Mexico 87545, USA}
\affiliation{$^2$ Ames National Laboratory, Ames, Iowa 50011, USA}
\begin{abstract}
	
\end{abstract}
\date{\today}
\maketitle

\date{\today}

1. In the Comment by Gurevich and Vinokur \cite{comm} and in Refs.~\onlinecite{BGK,BGBK,GV}, the vortex crossing rate was calculated in the framework of the London theory
of superconductivity treating vortex as a particle.
It allows one to derive the energy of a single vortex in thin-film superconducting strips as presented in Eq.\,(1) of the Comment.
However, one drawback of the London theory is that one needs to introduce a cutoff $\xi_{cut}$, due to the size of the vortex core, of the order of the 
coherence length $\xi(T)$ at the temperature $T$. 
Since the London theory treats vortices as pointlike objects with a phase, Eq.\,(1) in the Comment
is not valid   at distances smaller than $\xi(T)$ from the strip edges. The cutoff, an inherent short-coming of the  model, cannot be fixed within the London theory.
On the other hand, the vortex crossing rate, $R_v(I,T)$,  is very sensitive to the value of $\xi_{cut}$:\cite{BGBK} 
\begin{equation}
R_v(I,T) \approx \Omega(I,T, \nu) \left(\frac{e\pi I}{2 I_0}\right)^{\nu+1}
\left(\frac{\xi_{cut}}{w}\right)^{\nu+1},
\end{equation}
where $\Omega(I,T,\nu) \approx 4\pi Tc^2{R_{\rm eff}}/(e^2 \Phi_0^2)(\nu/2\pi)^{1/2}$ is the attempt frequency.
The characteristic current is $I_0={c\Phi_0}/({8\pi\Lambda})$, $w$ is the film's width,
${R_{\rm eff}}$ is the effective resistance and $e\simeq 2.718$. The exponent $\nu \approx 110$ was obtained in Ref.~\onlinecite{BGBK} from 
measurements of $R_v(I,T)$ on NbN.\cite{Bartolf}  Thus a change in $\xi_{cut}$ by 10\% results in a large change of $R_v(I,T)$ by the
factor $3\times 10^4$.
In Refs.~\onlinecite{BGK} and \onlinecite{GV} different values for $\xi_{cut}$ were used, thus resulting in very different crossing rates.  
We stress that without employing a microscopic theory
neither the London, nor the Ginzburg-Landau theories are sufficient to obtain   $\xi_{cut}$ accurately   
 at temperatures well below $T_c$, where experiments of interest are performed. 
  
In this situation, we can only  estimate the cutoff parameter $\xi_{cut}$ from the data by Bartolf {\it et al.}\cite{Bartolf} for $R_v(I,T)$. 
We obtain $\xi_{cut}(5.5 {\rm K})\approx 3.9$\,nm for sample 1 ($T_c=12.73$\,K). This  value should be compared with $\xi (5.5{\rm  K})\approx 3.2$\,nm  estimated from $H_{c2}$ measurements, see Fig.\,3 of Ref.\,\onlinecite{Bartolf}. For sample 2 the results are similar. On the other hand, according to Gurevich and Vinokur,   $\xi_{cut} (5.5{\rm K})\approx 3.9 \, {\rm nm}/(2\times 0.34)= 5.7$\,nm. 
We do not think that ``our" cutoff is any better justified than ``theirs".
As argued above, the discussion about the proper definition of $\xi_{cut}$ cannot be settled within the framework of the London theory. 

2. The   concept  of ``vortex as a particle" is a rather crude approximation to describe the vortex energy near the film edges.
It is questionable for the  process of vortex entry and exit, because such processes occur on the length scale of $\xi(T)$. 
In our view, only a microscopic theory can provide  the correct 
description. We do not think that the ``vortex as a particle" model can be improved significantly 
irrespective of the boundary conditions used.
Moreover, we are not aware of a convincing argument in favor of the periodic boundary conditions preferred by the authors of the Comment. Hence, we do not think that the corrections, presented in the Comment, improve this situation.
To put this all in perspective, one should keep in mind that the uncertainty in the factor $(\xi_{cut}/w)^{110}$ is 
much bigger than any corrections to the attempt frequency $\Omega$, due to a particular choice of boundary conditions.

3. In Refs.\,\onlinecite{BGK} and \onlinecite{GV} the single vortex crossing rate was derived for uncorrelated crossings along the strip length $L$. Vortices in the strip  are noninteracting (uncorrelated) at distances bigger
than  $w$. That is the reason why we evaluated the number of uncorrelated crossings of single vortices as $L/w$.

\end{document}